\documentclass[12pt,preprint]{aastex}

\usepackage{graphicx}

\newcommand{\bc}{\begin{center}}
\newcommand{\ec}{\end{center}}

\newcommand{\ie}{\textit{i.e.,\ }}
\newcommand{\eg}{\textit{e.g.,\ }}

\newcommand{\hst}{\textit{HST\ }}

\newcommand{\stwo}{\ensuremath{\mathrm{S_{\rm 2}}}}

\newcommand{\cstwo}{\ensuremath{\mathrm{CS_{\rm 2}}}}
\newcommand{\cotwo}{\ensuremath{\mathrm{CO_{\rm 2}}}}

\newcommand{\qco}{\ensuremath{\mathrm{Q_{\rm CO}}}}

\newcommand{\qwater}{Q\ensuremath{_\mathrm{H_{\rm 2}O}}}

\newcommand{\water}{\ensuremath{\mathrm{H_{\rm 2}O}}}

\newcommand{\formaldehyde}{\ensuremath{\mathrm{H_{2}CO}}}

\newcommand{\cf}{cf.,}
\newcommand{\api}{\ensuremath{\mathrm{a^{3}\Pi}}}
\newcommand{\xsigma}{\ensuremath{\mathrm{X^{1}\Sigma}}}
\newcommand{\asigma}{\ensuremath{\mathrm{A^{1}\Sigma}}}

\newcommand{\stardust}{\emph{Stardust}\ }

\newcommand{\epoxi}{\emph{EPOXI}\ }

\slugcomment{Accepted for publication in \emph{Astrophysical Journal Letters}}

\shorttitle{CO Abundance in 103P/Hartley}
\shortauthors{Weaver et al.}

\begin{document}


\title{The Carbon Monoxide Abundance in Comet 103P/Hartley\\ 
during the \emph{EPOXI} Flyby
}


\author{H. A. Weaver\altaffilmark{1},
P. D. Feldman\altaffilmark{2},
M. F. A'Hearn\altaffilmark{3},
N. Dello Russo\altaffilmark{1},
S. A. Stern\altaffilmark{4}
}

\altaffiltext{1}{Space Department,
Johns Hopkins University Applied Physics Laboratory,
11100 Johns Hopkins Road,
Laurel, MD 20723-6099;
hal.weaver@jhuapl.edu, neil.dello.russo@jhuapl.edu}

\altaffiltext{2}{Department of Physics and Astronomy,
The Johns Hopkins University,
3400 N. Charles Street,
Baltimore, Maryland 21218;
pdf@pha.jhu.edu}

\altaffiltext{3}{Department of Astronomy,
University of Maryland,
College Park, MD 20742;
ma@astro.umd.edu}

\altaffiltext{4}{Southwest Research Institute, 
Boulder, CO 80302;
alan@boulder.swri.edu}




\begin{abstract}
We report the detection of several emission bands in the CO Fourth Positive Group from 
comet 103P/Hartley during ultraviolet spectroscopic observations from the 
\emph{Hubble Space Telescope (HST)} on 2010 November~4 near the time of 
closest approach by NASA's \emph{EPOXI} spacecraft. 
The derived CO/\water\ ratio is 0.15--0.45\%, which places 103P 
among the most CO--depleted comets. 
Apparently this highly volatile species, whose abundance varies by a factor of 
$\sim$50 among the comets observed to date, does not play a major role in 
producing the strong and temporally variable 
activity in 103P/Hartley.
The CO emissions varied by $\sim$30\% between our two sets of observations, apparently in 
phase with the temporal variability measured for several gases and dust by other observers. 
The low absolute abundance of CO in 103P suggests several possibilities: 
the nucleus formed in a region of the solar nebula that was 
depleted in CO or too warm to retain much CO ice, 
repeated passages through the inner solar system have 
substantially depleted the comet's primordial CO reservoir, 
or any CO still in the nucleus is buried below the regions
that contribute significantly to the coma.
\end{abstract}





\section{INTRODUCTION}

Comet 103P/Hartley (hereafter ``103P'') was discovered in 1984 and is a 
Jupiter family comet (JFC) with an orbital 
period of 6.4~yr. 
A succession of encounters with Jupiter in 1875 reduced the comet's perihelion distance 
(from 2.9~AU to 1~AU) and aphelion distance (from 8~AU to 6~AU) \citep{Carusi:1985}.
Thus, 103P appears to be a relative newcomer to the inner solar system and 
is one of the few comets to become an Earth crosser in the recent past.

During its 2010 apparition, 103P was only 0.12~AU from the Earth on 
October~21 and was generally well placed for remote observations for several
months near the time of its perihelion on October~28 when the Sun--comet distance was 1.08~AU.
In addition, the \epoxi spacecraft passed within 
700~km of the nucleus of 103P on 2010 November~4 \citep{Ahearn:2011}. 
For these reasons, we proposed a systematic ultraviolet (UV)
spectroscopic investigation of 103P with the \emph{Hubble Space Telescope (HST)}. 
Our team was allocated a total of 15 orbits of \hst time to observe 103P during the fall of 2010, 
which we divided into three 5--orbit observing runs: 
on September~25, on November~4, and on November~28. 
Here we report on our detection of carbon monoxide (CO) gas 
in the coma of 103P near the time of the \epoxi flyby on November~4.

\section{OBSERVATIONS AND DATA REDUCTION}

We used two different \hst instruments: the Cosmic Origins Spectrograph (COS) 
and the Space Telescope Imaging Spectrograph (STIS)
Different gratings were selected to target different species (Table~1). 
A 2\farcs5 diameter aperture, which projected to 
285~km at the comet on November~4, was used for all COS observations, and a
\mbox{$0\farcs2 \times 52\arcsec$} 
\mbox{(23~km $\times$ 5920~km)}
slit was used for the STIS observation. 

The \hst observations of comet 103P/Hartley were very challenging 
owing to the comet's relatively uncertain ephemeris and its fast 
apparent motion, \mbox{$\sim$280\arcsec\ hr$^{-1}$} relative to the background star field. 
A pointing offset maneuver of 7\farcs7 was uplinked and executed prior to each orbit on 
2010 November~4 to correct for the improvement in the comet's ephemeris between the 
times when the observations were planned and when they executed, several weeks later. 
\emph{HST's} pointing control system was programmed to follow the comet's apparent motion. 
During each orbit an onboard target acquisition procedure
located the comet's center of brightness, presumed to be centered on the nucleus, 
and then commanded the telescope to center the nucleus in the selected aperture. 
Analysis of the acquisition images showed that the autonomous process worked as 
designed and centered the nucleus in the aperture to an accuracy of $\sim$0\farcs05 for all observations.

No cometary molecular or atomic emissions were detected in the STIS spectra,
which will not be discussed further here.
For the COS data, we used extracted spectra from the ``x1dsum'' files produced by the COS calibration pipeline 
processing system (``CALCOS''). 
We also examined the flat-fielded spectral images (the ``flt'' files) to search for potential artifacts 
associated with the spectral extractions.
Further details on COS, its operations, 
its performance, and its data products can be found in
the COS Instrument Handbook\footnote{\url{http://www.stsci.edu/hst/cos/documents/handbooks/current/cos\_ihb.pdf}}
and in the COS Data Handbook\footnote{\url{http://www.stsci.edu/hst/cos/documents/handbooks/current/COS\_longdhb.pdf}}.

A spectrum of 103P taken with the COS/FUV configuration is displayed in Figure~1 and shows
the detection of CO emission in the Fourth Positive Group 
(4PG; \asigma -- \xsigma) near 1500~\AA.
This is the first measurement of CO in 103P and is the main result presented in this paper.
A spectrum of 103P taken with the COS/NUV configuration is displayed in Figure~2 and
was used to search for CO emission in the dipole-forbidden Cameron system 
\mbox{(\api -- \xsigma)} near 1900~\AA\ , which is a tracer for \cotwo\ in the comet.

\section{ANALYSIS AND DISCUSSION}

Given the importance of \cotwo\ for 103P \citep{Ahearn:2011}, 
we first briefly discuss our \cotwo\ result.
Although the \hst measurements cannot detect \cotwo\ directly, 
forbidden (\emph{prompt}) emission in the CO Cameron system 
is produced during the photodissociative excitation
of \cotwo\ in the coma and can be used infer the rate of \cotwo\ 
sublimation from the nucleus \citep{Weaver:1994cocam}. 
From our upper limit on emission in the (1,0) band (Figure~2), we place a 5$\sigma$ 
upper limit on the \cotwo\ production rate of 
\mbox{$2.0 \times 10^{27}$ molecules s$^{-1}$} 
(assuming an outflow velocity of 0.78~km~s$^{-1}$), 
which implies \mbox{\cotwo/\water\ $\leq$ 20\%} (the \water\ production rate is discussed below).
This upper limit is well above the \cotwo/\water\  abundance ratio of $\sim$7\% 
determined from \hst observations of CO Cameron emission in 1991 \citep{Weaver:1994cocam}
and $\sim$8\% determined from direct measurements of \cotwo\ emission at infrared wavelengths 
in 1998 by the \emph{Infrared Space Observatory} \citep{Colangeli:1999}. 
The \epoxi team directly observed infrared emission from both \cotwo\ and \water\ during 
the 2010 apparition \citep{Ahearn:2011}. 
For observations 55~hr prior to closest approach (CA) of the \epoxi spacecraft to the nucleus, 
they derive a \cotwo/\water\  ratio of 20\% near a time 
of ``maximum'' activity. 
From observations $\sim$2~days after CA, the \epoxi data showed that the 
\cotwo/\water\  ratio increased by a factor of  $\sim$2 as the activity level changed
from a minimum to a maximum, implying that 
\mbox{\cotwo/\water\  $\approx$ 10\%} near times of minimum activity. 
This latter value is only marginally larger than the \cotwo/\water\   ratios measured 
during the remote (non-{\it EPOXI}) observations of 103P during the 1991 and 1998 apparitions. 
Thus, it is not clear whether the \cotwo/\water\   ratio in 103P has significantly 
changed during the past two decades.

UV fluorescence of sunlight in the CO 4PG system was discovered during a sounding
rocket observation of comet \mbox{C/1975 V1-A (West)} in 1976 \citep{Feldman:1976} 
and has been used to measure the CO abundance in approximately a dozen 
comets since then \citep[cf.,][]{Bockelee:cometsII}.
We had previously searched unsuccessfully for CO 4PG emission 
from 103P during \hst observations in September 1991 
(using the Faint Object Spectrograph) and in January 1998 (using STIS). 
The more constrained upper limit from these measurements was 1\% (3$\sigma$) 
for the CO/\water\ abundance in 1991 \citep{Weaver:1994cocam}. 
With the higher sensitivity of COS, 
we were able to detect several bands of the CO 4PG system from 103P on November~4 (Figure~1), 
even though the \water\ production rate was $\sim$5 times lower in 2010 compared to 1991. 
Table~2 lists the detected CO 4PG bands with their wavelengths and integrated brightnesses. 
The CO production rate appears to have increased by $\sim$30\% between visits~6 and 9, 
which were separated in time by $\sim$14.5~hr.  
The atomic emissions from C and S showed a similar change, suggesting there 
was a global increase in cometary activity for visit~9 compared to visit~6.

Strong temporal variability correlated with the rotation of the nucleus was measured from 
multiple ground and space observatories. The rotational period of 103P was estimated to be 
$\sim$16.64~hr based on observations taken during 2009 \citep{Meech:2009b}. 
However, the rotational period increased to $\sim$18--19~hr and the rotation became complex 
(\ie not about a single axis) near the time of the \epoxi flyby \citep{Ahearn:2011,Drahus:2011,Meech:2011}. 
We used the extensive ground- and \emph{EPOXI}-based monitoring of 103P to assess 
the comet's level of activity during the \hst observations (Figure~3). 
The temporal changes recorded during the \hst observations seem to be consistent with the 
observed periodic variation in other species, which is apparently associated with the small 
end of the nucleus rotating in and out of sunlight \citep{Ahearn:2011}.

We employed a radiative transfer model for cometary CO 4PG emission \citep{Lupu:2007} 
to convert the observed aperture-averaged brightness to a CO column density. 
The model assumes the CO emission is produced by the fluorescence of CO molecules 
stimulated by sunlight (\ie solar fluorescence).
We used a solar ultraviolet spectrum appropriate for the current level of solar activity, 
and performed line-by-line calculations, assuming a rotational temperature of 75~K and 
accounting for solar spectral features.
Our best model match to the observed average 103P spectrum yields a CO column density of 
\mbox{$1.15 \times 10^{13}$ molecules cm$^{-2}$.}

We used a standard radial outflow model \citep[cf.,][]{Bockelee:cometsII} 
to convert the column density to a gas production rate at the nucleus, 
assuming spherically symmetric outflow with constant speed. 
For the high spatial resolution observations discussed here (1\arcsec\ projects to 114~km at the comet), 
the finite lifetime of the CO molecule \mbox{($\sim$$1.5 \times 10^{6}$ s)} 
can be ignored, and the gas production rate can be calculated from 
$Q = v d N$, where $Q$ is the gas production rate \mbox{(molecules s$^{-1}$),} 
$v$ is the gas outflow velocity \mbox{(cm s$^{-1}$),}
$d$ is the diameter of the circular aperture (in cm at the comet; 
\mbox{$2.83 \times 10^{7}$ cm} for the COS aperture on November~4), 
and $N$ is the aperture-averaged gas column density (see above). 
We adopt \mbox{$v = 7.8 \times 10^{4}$ cm s$^{-1}$,} 
which is an appropriate value for this moderately active comet \citep[cf.,][]{Bockelee:cometsII}. 
Thus, the average CO production rate from our observation on November~4 is 
\mbox{$2.6 \times 10^{25}$ molecules s$^{-1}$;} 
the value during visit ~6 was $\sim$15\% smaller 
\mbox{($2.2 \times 10^{25}$)} and the value during visit~9 was 
$\sim$15\% larger \mbox{($2.9 \times 10^{25}$).}

The statistical uncertainty in \qco, based on the measured signal-to-noise 
ratio in the average spectrum, is $\sim$20\% (1$\sigma$). 
However, there are several sources of systematic error. 
The absolute calibration uncertainty is $\sim$15\%. 
Based on previous results on similarly active comets, the estimated uncertainty 
in the outflow velocity is $\sim$25\%. 
Another source of error is the mismatch between our fluorescence model 
and the observed CO band intensities. 
For example, the model predicts that the (2,0) band is slightly brighter than the (1,0) band 
whereas the observations indicate the opposite.
However, this discrepancy can be attributed to a detector artifact, which
has $\sim$3$\times$ larger dark current than the surrounding region, near
the location of the (1,0) band. 
There are other discrepancies, of varying magnitudes (Figure~1, Table~2), between the measured and 
predicted CO band intensities. 
Taking all of these issues into account, we conclude that the true value of \qco\ could be up to 50\% 
different from our adopted value.

The relative abundance of CO in 103P is the ratio \qco/\qwater, 
where \qwater\ is the water production rate. 
Although \qwater\ was not measured during our \hst observations, 
an excellent record was obtained during ground-based infrared (IR) observations 
on November~4 \citep{Dello:2011}. 
\qwater\ \mbox{(molecules s$^{-1}$)} increased by a 
factor of $1.6$ from the first IR observation 
\mbox{($8.84 \pm 1.09 \times 10^{27}$,} 
\mbox{10:49--11:30 UT)} to the last one 
\mbox{($1.38 \pm 0.16 \times 10^{28}$,} \mbox{15:20--15:54 UT),} 
but \qwater\  appeared to be leveling off during the final 2.5~hr of the IR observations 
\mbox{($1.40 \pm 0.17 \times 10^{28}$,} \mbox{14:50--15:17 UT).} 
Our first CO observation was 5.5~hr before the first IR observation, 
and our second CO observation was 4.0~hr after the last IR observation. 
By comparing when the \hst and IR measurements were made relative to 
the periodic changes in the comet's overall level of activity (Figure~3), 
we find that \hst visit~6 was near the minimum activity level, and 
visit~9 was near the average activity level. 
The IR measurements were made during the rising portion of the activity phase curve, 
starting near the average level and ending at approximately the maximum. 
Assuming a peak-to-valley activity ratio of 2 for \water\ \citep{Ahearn:2011}, 
we therefore conclude that \qwater\ was \mbox{$7.0 \times 10^{27}$} 
during visit~6 and \mbox{$1.0 \times 10^{28}$} 
during visit~9, yielding CO abundances of $\sim$0.31\% and $\sim$0.29\%, respectively. 
Given the large uncertainties discussed above, we estimate that the CO abundance 
was 0.15--0.45\%.

The CO abundance in 103P is apparently the lowest ever measured in a comet. 
Comets display a huge variation in the CO abundance, 
ranging from $\sim$0.4\% to $\sim$30\%  \citep{Bockelee:cometsII}, 
but the reason why is unknown. 
Although the most CO--rich comets are long-period Oort cloud objects, 
some long period comets have very low CO abundances 
($\sim$0.4\% for C/2001 WM1 [LINEAR]).
There is not much information available on the CO abundance in Jupiter family 
comets (JFCs), the dynamical class to which 103P belongs, owing to their low activity levels.  
Infrared CO emission was detected from JFC 21P/Giacobini-Zinner in October 1999, 
implying a CO abundance as high as $\sim$15\% \citep{Mumma:2000gz}. 
But IR CO emission was not detected from 21P several weeks later, when the comet was brighter, 
giving a 5$\sigma$ CO abundance upper limit of $\sim$3\% \citep{Weaver:1999gz}. 
IR observations of JFC 73P/Schwassmann-Wachmann~3-C gave a CO abundance of 
\mbox{$0.50 \pm 0.13$\%} \citep{Disanti:2007sw3}, 
comparable to the highest values allowed for 103P. 
A CO abundance of $\sim$4--6\% was measured from JFC 9P/Tempel, 
both before and after the outburst associated with the 
\emph{Deep Impact} event \citep{Mumma:2005t1,Feldman:2006t1}. 
A radio detection of CO emission from JFC 17P/Holmes yielded a 
CO abundance of $\sim$2--4\% \citep{Biver:2008holmes}, 
but the observations were made shortly after the spectacular outburst of the 
nucleus in October 2008 and at a relatively large heliocentric distance, 
both of which are rather unusual circumstances. 
Similarly, it is difficult to assess the compositional implications of the detection of 
radio CO emission from 29P/Schwassmann-Wachmann~1 \citep{Crovisier:2009jfc}, 
whose orbit never brings it within 5.7~AU of the Sun.

We assume here that all the observed CO is coming directly from the nucleus. 
In some comets, a significant fraction of the CO observed in the coma is
from ``extended sources'' (\eg \formaldehyde, \cotwo; 
\mbox{\cf\  \citealt{Bockelee:cometsII}).} 
However, our high spatial resolution \hst observations 
strongly favor the ``native'' source sublimating directly from the nucleus. 
We calculate that the photodissociation of \cotwo\ and 
\formaldehyde\ would contribute $<$10\% of the observed CO, 
assuming their relative abundances are 20\% and 1\%, respectively.
 
Similarly, the \water\ production rates quoted here assume all the \water\ is coming from 
sublimation at the nucleus. 
The results from the \epoxi mission suggest that a substantial fraction of the \water\ in
 the coma is produced by the sublimation of icy grains ejected in \cotwo\ jets \citep{Ahearn:2011}. 
 Ground-based IR observations show that the \water\ spatial profile is more extended than 
 expected for a parent species \citep{Dello:2011}, suggesting at least one-third of the 
 observed \water\ could be coming from an extended source (Dello Russo, priv.\ comm.). 
 If the icy grains presumed to be the source of the extended \water\ emission have a different 
 CO/\water\ ratio than is present in the bulk nucleus, the CO/\water\ abundance reported 
 here might not reflect the true composition of the nucleus.

\section{IMPLICATIONS}

CO is predicted to be the dominant C-bearing gas in the outer parts of the solar nebula 
where comets formed \citep[cf.,][]{Fegley:1989}. 
However, CO is extremely volatile with a sublimation temperature of $\sim$20~K under 
solar nebula conditions \citep{Yamamoto:1983}. 
Thus, the CO content of ices incorporated into comets may have been very sensitive 
to the local conditions where the nucleus formed, or where the coldest part of the nucleus 
formed assuming there was significant radial mixing of material within the nebula, 
as suggested by the results from the \stardust mission \citep{Brownlee:2006}. 
The capture of CO might have been enhanced by incorporation into a water ice clathrate 
\citep{Lunine:1989,Marboeuf:2010} or by trapping in amorphous water ice \citep{Barnun:2003}. 
But even if they were formed with significant amounts of CO, JFCs can subsequently 
lose most of their original CO by solar heating during multiple passes through the inner solar system. 
This may not explain the CO depletion in 103P, which only recently had its perihelion distance substantially reduced.
Another possible explanation is that any CO remaining in the nucleus is now buried
below the regions that contribute significantly to the coma.
In any case, our \hst result suggests that CO is probably a minor player in the comet's current activity and 
is not responsible for ejecting icy grains into the coma.

\acknowledgments

Based on observations made with the NASA/ESA \emph{Hubble Space Telescope,}
with data obtained from the archive at the 
Space Telescope Science Institute (STScI). 
We thank Steven Chesley for outstanding ephemeris support. 
We gratefully acknowledge Alison Vick, Merle Reinhart, 
David Sahnow, Tracy Ellis, for their expert help 
for these very demanding observations. 
Financial support for our \hst program (GO--12312) was provided by NASA 
through a grant from the STScI, which is operated 
by the Association of Universities for Research in Astronomy, 
Incorporated, under NASA contract NAS5--26555.

{\it Facility:} \facility{HST (COS, STIS)}

\clearpage


\clearpage

\begin{deluxetable}{ccccccc}	
\tabletypesize{\scriptsize}
\tablecaption{Log of \emph{Hubble Space Telescope (HST)} spectroscopic observations of 
comet 103P/Hartley on 2010 November~4.}
\tablewidth{0pt}
\tablehead{
\colhead{Visit \#\tablenotemark{a}} & \colhead{Instrument/\tablenotemark{b}} 
& \colhead{Grating\tablenotemark{c}} & \colhead{Central $\lambda$\tablenotemark{d}} 
& \colhead{Start Time\tablenotemark{e}} & \colhead{Exposure\tablenotemark{f}}
& \colhead{Objective\tablenotemark{g}} \\	
\colhead{Rootname} & \colhead{Configuration}	& & \colhead{(\AA)} 
& \colhead{(UT HH:MM:SS)} & \colhead{Time (s)} 
&	
}
\startdata
6		& COS/FUV	& G160M	& 1589	& 05:16:26	& 1310	& CO 4PG\\
LBK606	&			&		&		&			&		& for CO\\
7		& COS/NUV	& G185M	& 1986	& 11:39:38	& 1364	& CO Cameron\\
LBK607	&			&		&		&			&		& for \cotwo\\
8		& STIS/CCD	& G230LB	& 2375	& 13:02:02	& 1800	& OH for \water,\\
LBK608	&			&		&		&			&		& CS for \cstwo, \stwo\\
9		& COS/FUV	& G160M	& 1589	& 19:43:51	& 1100	& CO 4PG\\
LBK609	&			&		&		&			&		& for CO\\
10		& COS/NUV	& G185M	& 1986	& 22:59:14	& 866	& CO Cameron\\
LBK610	&			&		&		&			&		& for \cotwo\\

\enddata
\tablenotetext{a}{The Visit \# and Rootname refer to the values assigned by the STScI.}
\tablenotetext{b}{The Cosmic Origins Spectrograph (COS) was used in either its far ultraviolet (FUV)
or near ultraviolet (NUV) configuration.The Space Telescope Imaging Spectrograph (STIS) was 
used in its CCD configuration.}
\tablenotetext{c}{The grating designations refer to the values assigned by the STScI.}
\tablenotetext{d}{The central wavelengths ($\lambda$) for each grating are listed in angstroms (\AA).}
\tablenotetext{e}{The Start Times are universal time (UT) values at the Earth; the time of closest 
approach of the \epoxi spacecraft was 13:59:47 UT.}
\tablenotetext{f}{The exposure time is the total integration time in seconds (s) and was typically
divided among several individual exposures.}
\tablenotetext{g}{The Objective refers to the specific cometary emissions and molecules being targeted; 
see the text for further discussion.}
\tablecomments{The comet's geocentric and heliocentric distances were 0.157~AU and 1.064~AU, respectively, 
the Sun-Comet-Earth angle was 58\fdg8, and the Sun-Earth-Comet angle was 113\degr. 
Only the data from Visit \#s 6 and 9 are discussed in detail in this paper.}
\end{deluxetable}

\clearpage

\begin{deluxetable}{cccccc}
\tablecaption{Emissions in the CO Fourth Positive Group detected during  \emph{Hubble Space Telescope (HST)}
spectroscopic observations of  comet 103P/Hartley on 2010 November~4.}
\tablewidth{0pt}
\tablehead{
\colhead{Band\tablenotemark{a}} & \colhead{Wavelength\tablenotemark{b}} 
&\multicolumn{3}{c}{Brightness (Rayleighs)\tablenotemark{c}} & \colhead{Model\tablenotemark{d}}\\
\colhead{(v$'$,v$''$)} & \colhead{(\AA)} 
& \colhead{V6} 		& \colhead{V9}		& \colhead{Average}		& \colhead{(Rayleighs)}
}
\startdata
(4,0)		& 1419.10		& $0.22 \pm 0.19$	& $1.00 \pm 0.28$	& $0.61 \pm 0.17$	& $0.39$\\
(3,0)		& 1447.40		& $0.58 \pm 0.30$	& $0.61 \pm 0.30$	& $0.59 \pm 0.21$	& $1.16$\\
(2,0)		& 1477.60		& $0.84 \pm 0.40$	& $1.67 \pm 0.44$	& $1.25 \pm 0.29$	& $1.69$\\
(1,0)		& 1509.90		& $1.68 \pm 0.42$	& $2.43 \pm 0.51$	& $2.06 \pm 0.33$	& $1.55$\\
(3,2)$+$(0,0)	& 1542.51$+$1544.46 & $0.38 \pm 0.54$ & $3.02 \pm 0.72$ & $1.70 \pm 0.44$ & $0.95$\\
\enddata
\tablenotetext{a}{The CO bands are designated by their upper (v$'$) and lower (v$''$) vibrational levels.}
\tablenotetext{b}{The wavelengths are for the band heads; the bands are degraded to
the red (\ie extend asymmetrically to longer wavelengths).}
\tablenotetext{c}{The brightness is the band-integrated, aperture-averaged column brightness,
and its 1$\sigma$ error, for the visit~6 (V6), visit~9 (V9), and average spectra displayed
in Figure~1 \mbox{(1 Rayleigh $= 10^{6}/4\pi$ photons cm$^{-2}$ s$^{-1}$ sr$^{-1}$)}.}
\tablenotetext{d}{The model of \citet{Lupu:2007} is matched to the observed spectra.}
\end{deluxetable}

\clearpage

\begin{figure}
\epsscale{1.}
\begin{center}
\plotone{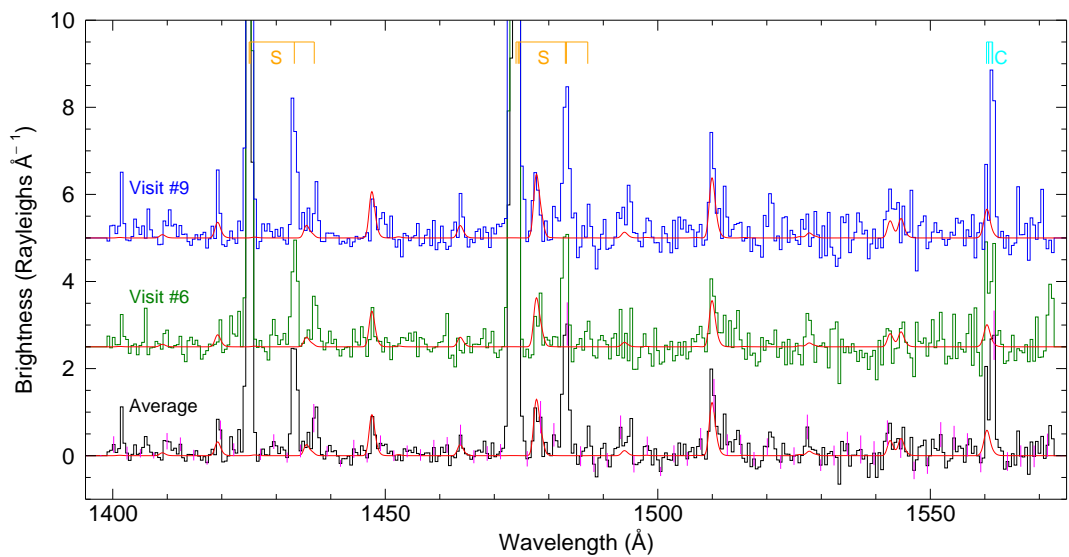}
\caption{Portions of the COS/FUV G160M spectra 
of 103P/Hartley from visits~6 (green), 9 (blue), and their unweighted average (black) are displayed.
The original data were re-binned by a factor of 40 to improve the signal-to-noise ratio. 
Statistical errors are shown (magenta) for every fifth point in the average spectrum.
The spectra for visits~6 and 9 have been shifted up by 2.5 and 5 y--units, respectively, for clarity. 
The strongest observed emissions are from sulfur (S) and carbon (C) atoms in the 
coma, but weak CO 4PG emission is also detected (Table~2). 
A model \citep{Lupu:2007} fit for the CO emission is plotted in red 
for all displayed spectra. 
 \label{fig:co4pg} } 
\end{center} 
\end{figure}

\begin{figure}
\epsscale{1.}
\begin{center}
\plotone{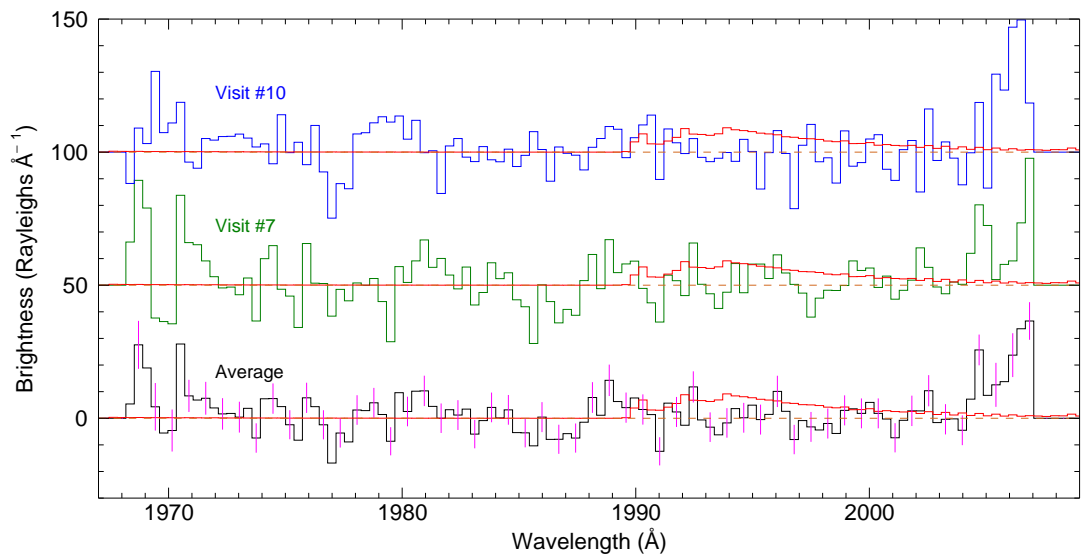}
\caption{Portions of the COS/NUV G185M spectra of comet 103P/Hartley
from visits~7 (green), 10 (blue), and a weighted average (black)
are displayed for a region covering the strongest bands of the 
CO Cameron system.
The original data were re-binned by a factor of 10 to improve the signal-to-noise ratio. 
Statistical errors are shown (magenta) for every other point in the average spectrum. 
The spectra for visits~7 and 10 have been shifted up by 50 and 100 
y-units, respectively, for clarity. 
No Cameron band emission was detected. 
A model for the CO Cameron band emission \citep{Conway:1981} 
overlays each of the spectra (red) and corresponds approximately
to a 5$\sigma$ upper limit on the level of potential cometary emission. 
\label{fig:cocam} } 
\end{center} 
\end{figure}

\begin{figure}
\epsscale{1.}
\begin{center}
\plotone{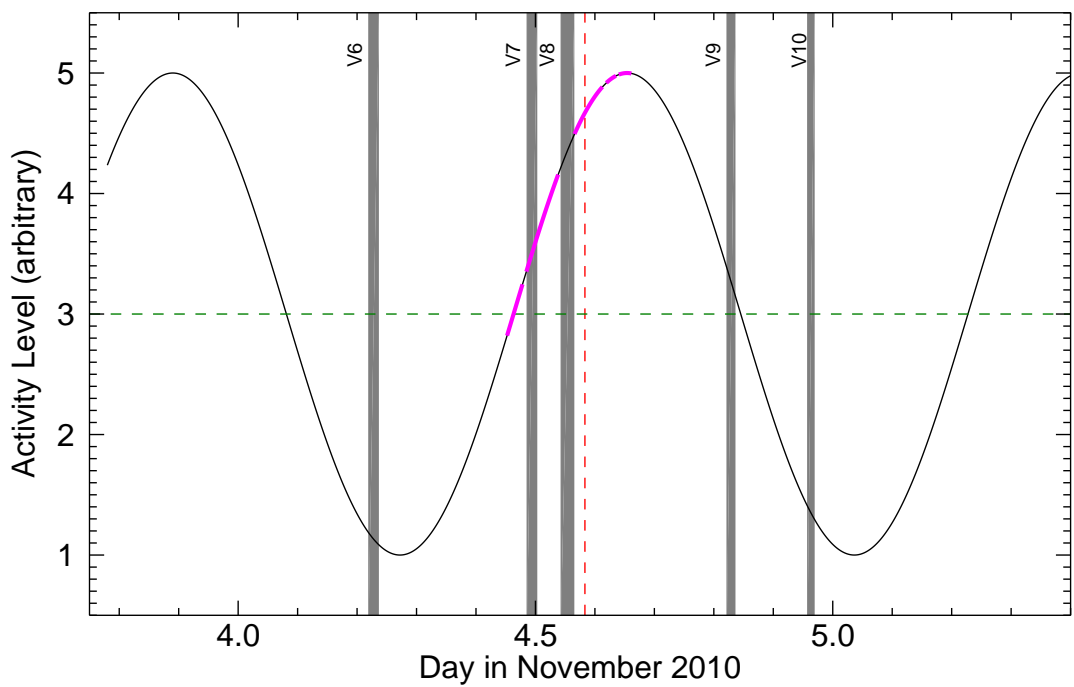}
\caption{This is a schematic depiction of the  \emph{relative} activity level of 
comet 103P/Hartley near the time of the \epoxi closest approach (CA) on 2010 November~4.5832 UT.
The solid curve is a pure sinusoid with a period of 18.34~hr, as derived from visible
light curve data from \epoxi \citep{Ahearn:2011} and from ground-based HCN mm-wave lineshape 
data \citep{Drahus:2011}.
The vertical scale is in arbitrary units but should accurately represent the times when the comet's 
activity was at a local minimum or maximum, or somewhere in-between. 
The red vertical dashed line shows the \epoxi spacecraft CA time. 
The time windows for the 5 \hst visits (V6, V7, V8, V9, V10; see Table~1) 
are depicted by the gray rectangles. 
The magenta-colored curves show the 5 time windows for the infrared measurements 
of various gas production rates \citep{Dello:2011}. 
\label{fig:phase} } 
\end{center} 
\end{figure}

\end{document}